 \documentclass[final,5p,times,twocolumn,authoryear]{elsarticle}

\usepackage{amssymb}
\usepackage{lipsum}

\journal{Astronomy $\&$ Computing}

\begin{document}

\begin{frontmatter}



\title{SVOM Follow-up Observation Coordinating Service}


\author[aff1,aff2]{Xu-hui Han}
\ead{hxh@nao.cas.cn}
\author[aff1]{Pin-pin Zhang\corref{cor1}}
\ead{ppzhang@nao.cas.cn}
\author[aff1]{Yu-jie Xiao}
\ead{yjxiao@bao.ac.cn}
\author[aff1]{Ruo-song Zhang}
\author[aff1,aff2]{Chao Wu}
\author[aff1,aff2]{Li-ping Xin}
\author[aff1]{Hong-bo Cai}
\author[aff3]{Hai Cao}
\author[aff4]{Hui-jun Chen}
\author[aff1,aff2]{Jin-song Deng}
\author[aff1]{Wen-long Dong}
\author[aff5]{Guo-wang Du}
\author[aff1]{Lei Huang}
\author[aff1]{Lin Lan}
\author[aff1]{Hua-li Li}
\author[aff1,aff2]{Guang-wei Li}
\author[aff1]{Xiao-meng Lu}
\author[aff1]{Yu-lei Qiu}
\author[aff6]{Jian-feng Tian}
\author[aff1,aff2]{Jing Wang}
\author[aff1]{Wen-jin Xie}
\author[aff1,aff2]{Da-wei Xu}
\author[aff1]{Yang Xu}
\author[aff1]{Zhu-heng Yao}
\author[aff1]{Xue-ying Zhao}
\author[aff6]{Jie Zheng}
\author[aff7]{Wei-kang Zheng}
\author[aff1]{Ya-tong Zheng}
\author[aff8]{Xiao-xiao Zhou}
\author[aff1,aff2]{Jian-yan Wei}
\affiliation[aff1]{organization={CAS Key Laboratory of Space Astronomy and Technology, National Astronomical Observatories, Chinese Academy of Sciences},
            city={Beijing},
            postcode={100101}, 
            country={People's Republic of China}}
\affiliation[aff2]{organization={School of Astronomy and Space Science, University of Chinese Academy of Sciences},
            city={Beijing},
            postcode={100101}, 
            country={People's Republic of China}}
\affiliation[aff3]{organization={School of Space Science and Technology, Shandong University,},
            city={Weihai, Shandong},
            postcode={264200}, 
            country={People's Republic of China}}
\affiliation[aff4]{organization={Department of Computer Science and Technology, School of Computer and Information Technology Cangzhou Jiaotong College},
            city={Cangzhou, Hebei},
            postcode={061199}, 
            country={People's Republic of China}}
\affiliation[aff5]{organization={South-Western Institute for Astronomy Research, Yunnan University},
            city={Kunming, Yunnan},
            postcode={650504}, 
            country={People's Republic of China}}           
\affiliation[aff6]{organization={National Astronomical Observatories, Chinese Academy of Sciences},
            city={Beijing},
            postcode={100101}, 
            country={People's Republic of China}}
\affiliation[aff7]{organization={Department of Astronomy, University of California},
            city={Berkeley},
            postcode={CA 94720-3411}, 
            country={USA}}
\affiliation[aff8]{organization={College of Information Engineering, Fuzhou Polytechnic},
            city={Fuzhou, Fujian},
            postcode={350108}, 
            country={People's Republic of China}}
\begin{abstract}
The Sino-French SVOM (Space Variable Objects Monitor) mission is a space-based astronomy mission complemented with ground-based dedicated instrumentation. It 
aims to explore and study high-energy cosmic phenomena, such as gamma-ray bursts (GRBs). This unprecedented combination of space-based and ground-based 
instruments will provide leading multi-wavelength observational capabilities in gamma-rays, X-rays, optical, and near-infrared bands. The complete observation 
sequence of each GRB triggered by the SVOM mission consists of three stages, the GRB detections, followed by the on-board and grounded automatic follow-ups, and rapid 
deep multi-band photometry and spectroscopy re-visit observations. To efficiently organize all grounded instruments performing automatic follow-ups and re-visit observations, 
we develop a follow-up observation coordinating service (FOCS), which is capable of performing GRB trigger distributing, automatic observation scheduling and observation 
coordination supporting by providing a user support platform. The FOCS also facilitates the provision of observational planning for ground-based telescopes to conduct synchronized observations of identical celestial regions as SVOM. The FOCS is utilized for the SVOM-dedicated ground-based telescopes as well as for associated partner telescopes. Since the launch of SVOM in June 2024, as the FOCS system joining the operations of SVOM, multiple successful 
observations have been made for SVOM GRBs. In this paper, we present the goals of the FOCS system as well as the principle and workflow developed to achieve these 
goals. The structure, technical design, implementation, and performance of the FOCS system are also described in detail. We conclude with a summary of the current 
status of the FOCS system and our near-future development plan.

\end{abstract}




\end{frontmatter}




\section{Introduction}
\label{introduction}

The Sino-French Space Variable Objects Monitor (SVOM) mission, launched in June 2024, marks a significant milestone in international collaborations between 
China and France in the field of space exploration. The primary objective of SVOM is to monitor variable objects in space, with a particular focus on gamma-ray 
bursts (GRBs), which are among the most powerful explosions in the universe \citep{Atteia22}. By studying GRBs, scientists aim to gain deeper insights into 
the formation and evolution of stars, galaxies, and potentially the early universe itself.
SVOM comprises both space-based and ground-based components. The satellite is equipped with several instruments, including GRM (Gamma Ray Burst Monitor), 
ECLAIRs (an X-ray and low-energy gamma-ray telescope), MXT (Microchannel X-ray Telescope) and VT (Visible Telescope). Observations from space will be 
complemented by a large ground segment consisting of wide-field cameras (GWAC - Ground-based Wide Angle Camera), robotic telescopes (GFTs - Ground Follow-up 
Telescopes), and photometric and spectroscopic telescopes. These instruments work in concert to provide multi-wavelength observations, allowing researchers 
to capture detailed information about GRBs and other cosmic phenomena. The scientific goals of SVOM include not only the detection of new GRBs but also the 
study of their physical properties, such as energy spectra, light curves, and host environments.

The observation of GRB triggered by the SVOM satellite system consists of three stages.

\textbf{Stage 1: Initial Detection}

ECLAIRs and GRM detect the GRBs, while the ground-based GWAC simultaneously captures the optical transient radiation.

\textbf{Stage 2: Rapid Follow-up}

Upon triggering of the GRBs, the satellite autonomously reorients itself, and the onboard MXT and VT conduct automatic observations within 5 minutes. 
Simultaneously, within 1-2 minutes, the GRB trigger information is transmitted to ground-based telescopes via VHF and BeiDou short messages. 
The SVOM-dedicated follow-up telescopes, CGFT and FGFT, initiate rapid automatic follow-up observations. Additionally, some associated partner 
telescopes within the SVOM global ground-based follow-up telescope network will also conduct rapid-response to do automatic follow-up observations.

\textbf{Stage 3: Deep Analysis } 

Based on the data analysis results from the first and second stages, GRB candidates with significant research value are quickly screened, and further observations, 
known as GRB re-visit observations, are organized. These re-visit observations include deep multi-band photometry and spectroscopic observations, thus enable 
astronomers to identify the host galaxy of a GRB, providing clues about the conditions under which the burst occurred. For instance, observing the spectral 
lines of the host galaxy can reveal its redshift, indicating its distance and offering insights into the epoch of star formation. Given that some GRBs are 
thought to be associated with the collapse of massive stars or the merger of neutron stars, these observations are pivotal for understanding stellar evolution 
and gravitational wave astronomy.

The observations in the first and second stages are designed to be automatic. Unlike the satellite, the automated follow-up observations of ground-based telescopes 
from both China and France are managed separately by each party’s ground-based observation systems (this paper focuses solely on the Chinese ground-based 
observation system). To accommodate the characteristics of Chinese telescopes, technical support is provided for their automated follow-up observations. 
The observations in the third stage, however, are initiated manually after analyzing and judging the data from the previous two stages. Therefore, it is 
essential to prioritize the allocation of scarce telescope observation resources to GRBs with high scientific value. Consequently, establishing connections 
between scientists and telescopes and providing a user-friendly platform for streamlined access and coordination is crucial. 

Through the observations conducted in these three stages, a data foundation is established to integrate and analyze multi-wavelength observational data of GRBs, 
thereby enhancing the scientific output value of GRB research.

To streamline the entire process of three-stage observations, we developed a Follow-Up Observation Coordinating Service (FOCS) to maximize the likelihood 
and efficiency of successful observations using the SVOM Chinese ground-based follow-up telescopes. The FOCS workflow begins with the coordinated observation 
planning between ground-based telescopes and the SVOM satellite. Upon receiving a GRB trigger, FOCS rapidly distributes alerts to all participating ground-based 
instruments. This real-time notification enables observatories to prepare for and initiate observations swiftly, minimizing critical response time—a necessity 
given the transient nature of GRBs.

Another key objective is automating observation scheduling. Due to differing technical capabilities among telescopes, some cannot perform fully automated 
follow-up or re-visit observations, making manual scheduling impractical and inefficient. The FOCS automatically prioritizes GRB notices following 
strategies pre-defined by the users based on scientific value, observational urgency, and instrument-specific capabilities. This approach optimizes 
telescope resource utilization.

Furthermore, the FOCS system simplifies the process of conducting re-visit observations with larger telescopes. By providing a dedicated user support 
platform, it enables seamless integration between requesting and executing re-visit observations. This integrated workflow allows scientists to 
leverage telescopes more effectively.

\begin{figure}
	\centering 
	\includegraphics[width=0.5\textwidth]{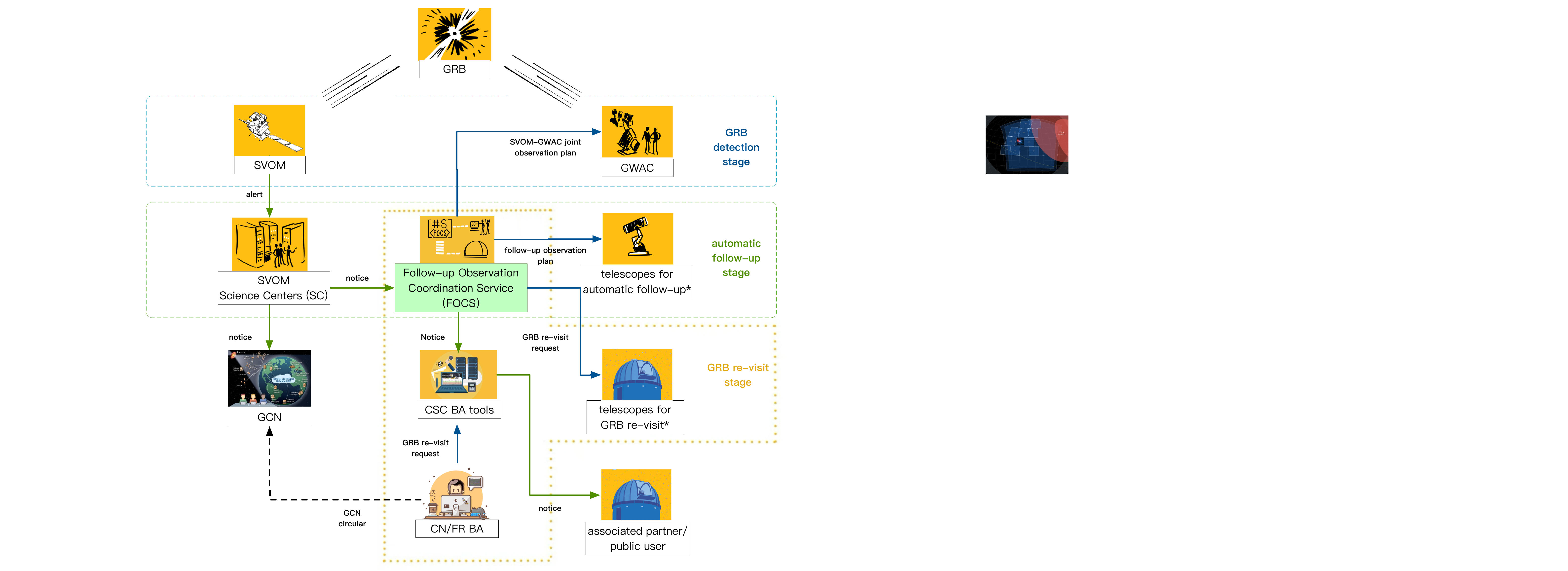}	
	\caption{The sequence of SVOM three-stage observations is shown in the figure, which illustrates the role of 
    FOCS in the observation sequence.} 
	\label{fig_svom_alert_followup}
\end{figure}

Therefore, the FOCS plays a critically important role in the Chinese ground-based observation system. Figure \ref{fig_svom_alert_followup} illustrates 
its role in the sequence of three-stage observations.
In the following sections of this paper, we will discribe the FOCS Architecture design, its implementation, performance and prospect of the future updates. 

\section{FOCS Architecture and Key Features}
\label{FOCS_Architecture}

In the system design process, we addressed three critical contradictions: First, to ensure scientific observational efficiency, the system must balance the 
detection probability of high-value targets (e.g., GRB optical counterparts) with optimal allocation of limited observational resources, imposing stringent 
requirements on system reliability and software robustness. Second, to accommodate the heterogeneous characteristics of ground-based telescope operation 
control systems (OCSs) across Chinese collaborating institutions, the system must possess cross-platform compatibility and technological inclusiveness. 
Third, while fulfilling the requirements for SVOM space-ground coordination and automated follow-up observations of GRBs, the system must preserve the 
autonomous decision-making authority of individual telescopes for their independent scientific missions, necessitating architectural flexibility with 
configurable strategies and scalable functionalities. Based on these considerations, FOCS established core design principles emphasizing reliability, 
universality, and adaptability.

The system adopts a Client-Server (C/S) architecture to support multi-telescope coordinated observations. The server is deployed at the SVOM Chinese Science Center (CSC), 
responsible for real-time reception and subscription-based distribution of GRB alerts. To ensure universality, standardized clients are constructed for each 
collaborating telescope to reliably transmit three critical data streams: 1) GRB alert information streams; 2) coordinated observation request streams; and 3) 
follow-up observation plan instruction streams for GRBs. Clients simultaneously establish bidirectional feedback channels to report message reception status 
and plan execution progress in real time, forming a complete closed-loop control system.

In terms of observation planning mechanisms, FOCS innovatively implements a hybrid planning model to accommodate diverse telescope scheduling strategies. 
For devices under centralized scheduling, the Centralized Planning Mode (CPM) is adopted, where the server generates globally optimized observation plans using 
multi-device optimization algorithms, enhancing the efficiency of coordinated multi-telescope observations. For autonomous telescopes, the Decentralized Planning 
Mode (DPM) is activated, enabling local observation decisions through client-side planning module. This design allows telescopes to independently resolve priority 
conflicts between SVOM missions and other scientific tasks. Notably, the system supports flexible decoupling of planning modules, permitting mature observation 
systems to directly utilize FOCS alert data for independent plan generation.

At the communication protocol level, a hybrid architecture balances system performance and reliability. The server communicates with CSC infrastructure using MQTT 
for alert data transmission, leveraging the publish/subscribe paradigm to ensure low-latency, high-reliability message distribution. Between server and clients, 
a MQTT+HTTP/2 protocol combination is employed: MQTT handles real-time alert data and observation plans/requests, while HTTP/2 ensures reliable transmission of 
large observational datasets (e.g., follow-up observation images).

The client output customization mechanism reduces integration barriers for telescopes. FOCS employs a configurable output architecture featuring:

\begin{itemize}
    \item A template library for the Observation Target Description Language (OTDL) that enables output parameter selection based on parameter classes;
    \item Configuration files defining parameter mapping rules to convert formats for telescope OCSs;
    \item Customized output configurations tailored to specific telescope control requirements.
\end{itemize}

This modular design with standardized interfaces ensures system reliability while achieving broad compatibility with heterogeneous observational terminals. 
Empirical testing demonstrates that new telescopes require as few as 8 person-hours for system integration.

This hierarchical architecture significantly enhances engineering applicability. FOCS has successfully integrated four categories of heterogeneous telescopes. 
The stability and reliability of FOCS have been validated through nearly nine months of operational testing following the SVOM launch (June 2024 to March 2025).

\begin{figure*}
	\centering 
	\includegraphics[width=0.8\textwidth]{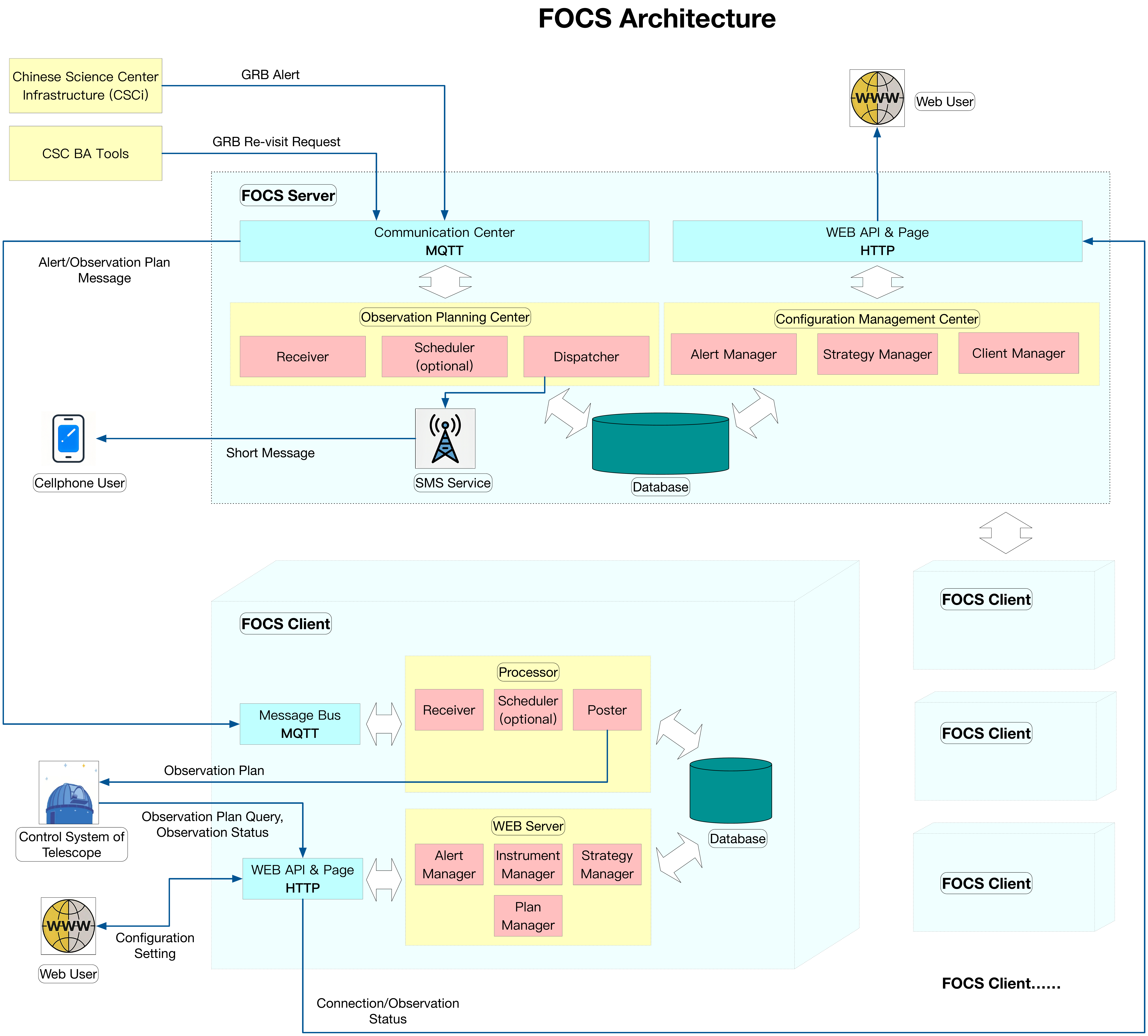}	
	\caption{The architecture of FOCS. The top square shows the architecture of server. The boxes indicate the clients.} 
	\label{fig_focs_architecture}
\end{figure*}

The foundational architecture of the system is depicted in Figure \ref{fig_focs_architecture}. The technical rationale behind protocol and architecture selections 
will be systematically elaborated in subsequent sections.

\section{Functional Module Design of the FOCS Server System}

The FOCS employs a distributed C/S architecture with elastic deployment capabilities, designed to achieve efficient and flexible GRB follow-up observations. 
The system consists of a logically decoupled centralized server and scalable clients, which interact via standardized Application Programming Interfaces (APIs) to 
enable data interoperability between heterogeneous systems. Clients can be dynamically added or removed, featuring plug-and-play functionality.

The server is deployed at the CSC of SVOM and provides the following core services:
\begin{itemize}
    \item \textbf{Automated observation planning and scheduling for ground-based telescopes}: 
    Automatically generates observation plans for specific ground-based telescopes based on SVOM satellite data and predefined strategies.
    \item \textbf{Distribution of SVOM GRB alerts and observation plans}: 
    Receives and disseminates SVOM GRB alert information and satellite observation plans.
    \item \textbf{Client node management and observation strategy configuration}: 
    Provides functionalities for client node management and supports the configuration and maintenance of observation strategies.
    \item \textbf{SMS alert service for SVOM GRB notifications}: 
    Offers users SMS notifications for SVOM GRB alert information.
\end{itemize}

To realize the aforementioned services, the server system comprises the following key functional modules:

\subsection{Communication Center}
Communication Center employs the MQTT (Message Queuing Telemetry Transport) protocol as its message bus. MQTT is a lightweight communication protocol based on a 
publish-subscribe model, suitable for resource-constrained environments.

The external inputs to the FOCS system primarily originate from three sources:

\begin{itemize}
    \item \textbf{SVOM satellite observation plans}: 
    Provided by the Chinese Science Center infrastructure (CSCi), containing information such as satellite pointing coordinates and rotation times.
    \item \textbf{SVOM GRB alerts}: 
    Also provided by the CSCi.
    \item \textbf{GRB re-visit observation requests}: 
    Generated by the Burst Advocate (BA) support tools of the CSC.
\end{itemize}

The FOCS server utilizes the MQTT message bus to distribute SVOM satellite observation plans, SVOM GRB alerts, and observation plans to clients. It 
also obtains feedback on client connection status and message reception status to ensure the reliability of message transmission.

\subsection{Observation Planning Center}

The observation planning center module is the central component of FOCS, responsible for receiving, parsing, standardizing GRB alerts generating, 
and distributing observation plans. This module encompasses the following three primary functions:

\begin{itemize}
    \item \textbf{Receiver} (Message reception, parsing, and standardization): 
    Receives various messages from the MQTT message bus, parses the message content, and standardizes it into a unified data format.
    \item \textbf{Scheduler} (Observation plan generation): 
    Generates optimized observation plans based on predefined observation strategies, SVOM satellite data, and the characteristics of ground-based telescopes.
    \item \textbf{Dispatcher} (Message distribution): 
    Distributes key parameters or observation plans from the messages to specified clients according to their subscription rules. Additionally, it sends key 
    parameters from alert information to designated users via SMS.
\end{itemize}

The observation plan generation process, upon receiving input messages, directs them into different processing branches based on the message category:

\begin{itemize}
    \item \textbf{SVOM satellite observation plan messages}: 
    Enter the “SVOM Satellite and Ground Telescope Collaborative Observation” processing branch. The parameter extraction module extracts information such as the observation 
    targets, observation modes, satellite pointing coordinates, and satellite rotation times from the observation plan, which is then input into the scheduling module. 
    The scheduling module checks the subscription needs of each client; if a client only requires the SVOM satellite observation plan, the parameter information is 
    directly sent to the client via the distribution module. If the client requires FOCS to complete a collaborative observation plan, the scheduling module reads the 
    geographical location information of the telescope and the sky region division information (large field telescopes have predefined fixed sky regions to cover the 
    entire celestial sphere), matches the observation regions of the SVOM onboard ECLAIRs telescope with the ground telescope regions, calculates the visibility and 
    time windows for the ground telescopes, and formulates the visible regions into ground telescope observation plans.

    \item \textbf{SVOM GRB alert messages}: 
    Enter the “GRB Automatic Follow-up Observation” processing branch. The parameter extraction module extracts information from the alert, including the GRB trigger number, 
    trigger time, coordinates, localization error range, and alert type, formats it consistently, and inputs it into the scheduling module. The scheduling module checks the 
    subscription needs of each client. If a client only requires SVOM GRB alert information, the parameter information is directly sent to the client via the distribution module. 
    If the client requires FOCS to complete a GRB automatic follow-up observation plan, the scheduling module invokes an independent GRB automatic follow-up observation plan 
    generation module (Alert2Target program), which generates standard ground telescope observation plans. The processing logic of the Alert2Target program will be elaborated 
    in subsequent sections.

    \item \textbf{GRB re-visit observation request messages}: 
    Enter the “GRB Re-visit Observation” processing branch. The parameter extraction module extracts information from the observation request, including the telescope 
    information used, observation targets, target positions, and observation parameter configurations, and inputs it into the scheduling module. The scheduling module matches 
    the used telescope with the corresponding client, formulates the observation request parameters into ground telescope observation plans, and sends them to the client via 
    the distribution module.
\end{itemize}

\begin{figure*}
	\centering 
	\includegraphics[width=0.8\textwidth]{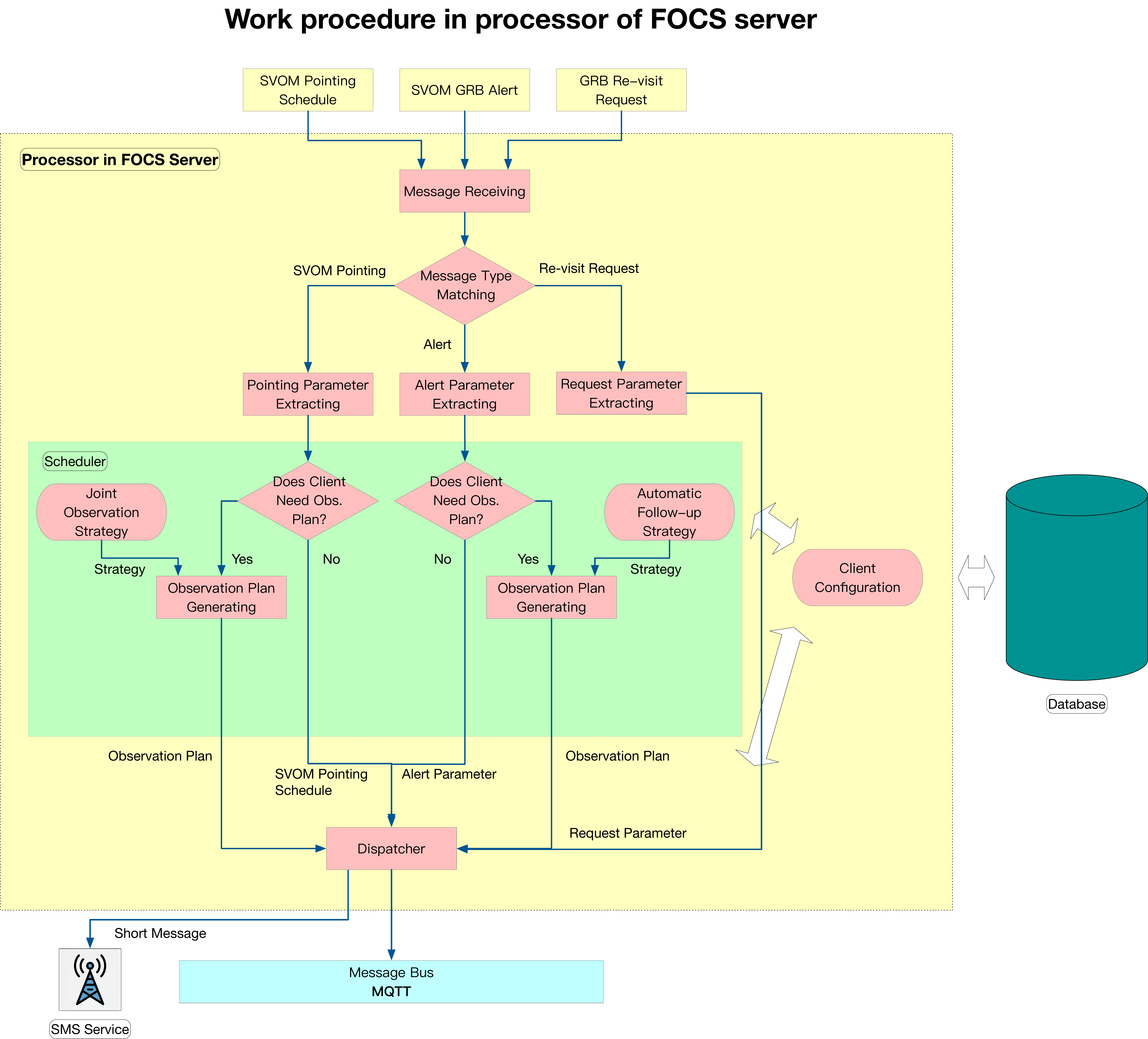}	
	\caption{The working procedure of the observation planning center of FOCS server is shown in the yellow box. The green square contents the scheduling strategies 
    for the SVOM and ground telescope collaborative observation and the GRB automatic follow-up observation. } 
	\label{fig_server_workflow}
\end{figure*}

These processing branches of the observation planning center are depicted in the Figure \ref{fig_server_workflow}, which also shows the working procedure of FOCS server. 

\subsection{Configuration Management Center}

The configuration management center module is developed based on a Web server, which employs a frontend-backend decoupled architecture. The frontend is built using HTML, CSS, and 
JavaScript, while the backend primarily leverages Java (SpringBoot framework) and Python for development.
The processing rules and logic of the core business processing module depend on pre-defined alert filtering conditions, sky region differentiation information, ground telescope 
pointing ranges, and observation strategies. These rules and strategies are defined based on the needs of client nodes, making the management of client nodes and these rules/strategies crucial.

To facilitate the management of alerts, observation strategies, and observation nodes, the server provides a cross-platform human-machine  interface (HMI) built on web technologies. 
This interface allows users to configure and manage settings in a visual environment. Additionally, the web page displays the connection status of nodes, as well as the distribution 
and reception status of alerts and messages, enabling users to intuitively view the operational status of each node.

\subsection{SMS Notification Service}
Considering the urgency of GRB follow-up observations, FOCS provides users with an SMS notification service for SVOM GRB alert information. The system groups users based on the 
categories of alerts they subscribe to, and the SMS service module sends specific types of alert parameters to users via SMS according to their subscription rules, ensuring that 
users can promptly receive critical information.

\section{Functional Module Design of the FOCS Client}

To address the challenges posed by the strong heterogeneity of technical foundations and the absence of a unified OCS among ground-based telescopes in the SVOM project, 
FOCS adopts an independent client architecture. This approach minimizes the complexity of software modifications required at the telescope end.

The client nodes integrate Browser-Server (B/S) architecture to implement core functional modules. Through multi-layered modularization and standardized interface 
design, the system achieves a highly decoupled structure, significantly enhancing scalability and continuous evolution capability. The client’s HMI, built on web 
technologies, enables flexible multi-terminal operation with zero-deployment advantages. Key client functionalities include:
(1) receiving SVOM GRB alert parameters, SVOM satellite observation plans, or observation schedules; (2) automatic observation planning; (3) observation plan format 
conversion and transmission; and (4) management of SVOM GRB alert configurations, instrument parameters, observation strategies, and observation plan rules. 
The client deployment architecture consists of three core modules:

\subsection{MQTT Message Bus Module}
This module acts as the asynchronous communication hub between the client and the server, utilizing the MQTT protocol for real-time transmission 
of multi-source observation information. It primarily handles three types of structured data:
\begin{itemize}
\item SVOM satellite observation plans (including satellite pointing coordinates and rotation timing parameters);
\item SVOM GRB alert parameters;
\item Composite observation plans (including satellite-ground collaborative observation plans, GRB automatic follow-up 
observation plans, and GRB re-visit observation plans).
\end{itemize}
This module incorporates a bidirectional status feedback mechanism, allowing real-time reporting of the client connection status and message reception confirmation, 
thereby ensuring the reliability of observation instruction transmission.

\subsection{Core Business Processing Module (Processor)}
The Processor module employs a multi-threaded asynchronous processing mechanism, comprising a three-level functional structure:

\begin{itemize}
    \item \textbf{Receiver} (Message routing): 
    Establishes a multi-channel processing mechanism based on message type, injecting input data into corresponding processing pipelines.
    \item \textbf{Scheduler} (Observation planning):
    \textit{Collaborative Observation Branch}: Reserves an interface for generating satellite-ground collaborative observation plans; currently not activated in 
    the current version.
    \textit{GRB Follow-up Observation Branch}: Integrates the Alert2Target program to automatically generate standard observation plans based on user-defined 
    strategies (including alert type filtering, signal-to-noise ratio thresholds, etc.). Its decision-making process includes: alert type 
    identification → strategy library matching → visibility analysis → exposure parameter binding → observation plan generation.
    \textit{Observation Plan Processing Branch}: Implements observation plan format conversion and transmission protocol adaptation.
    \item \textbf{Poster} (Message distributing):  Supports both active push (via WEB API) and passive query dual transmission modes,
    providing an extensible format conversion plugin system to accommodate the heterogeneous interface specifications of different telescope OCSs.
\end{itemize}

\subsection{Web Service Module}
This module adopts a frontend and backend separated architecture. The frontend is built on “Vue3 + Axios + Element Plus” to construct a responsive management 
interface, and the backend uses the Django framework to implement API services. The main functions include:

\begin{itemize}
    \item \textbf{Display of Multi-dimensional Data}: 
    Real-time display of GRB alert, observation plan execution status, and equipment operating parameters.

    \item \textbf{Strategy Configuration System}:
    
    \textit{Telescope Parameter Management}: 
    Supports geographic coordinate registration, field-of-view (FOV) parameter configuration, and equipment status monitoring.
    
    \textit{Observation Strategy Optimization}: 
    Provides hierarchical alert filtering conditions (delay, positioning accuracy, confidence level, etc.), visibility analysis tools (solar altitude 
    angle threshold, target horizon altitude constraints), and pointing decision algorithms (pointing necessity criteria based on FOV coverage).
    
    \textit{Exposure Template Management}: 
    Establishes a parameterized exposure combination library (band selection, exposure duration, imaging frame number, etc.), supporting strategy-driven 
    parameter combination calls.
    
\end{itemize}

The client is implemented in Python ($\geq$3.10) – a widely adopted and developer-friendly language – to ensure compliance with functional requirements while enhancing development, deployment, and integration efficiency. Through modular architecture and open interface design, it streamlines interactions with OCSs, effectively addressing integration challenges caused by the heterogeneous nature of ground-based telescope systems.

\begin{figure*}
	\centering 
	\includegraphics[width=0.8\textwidth]{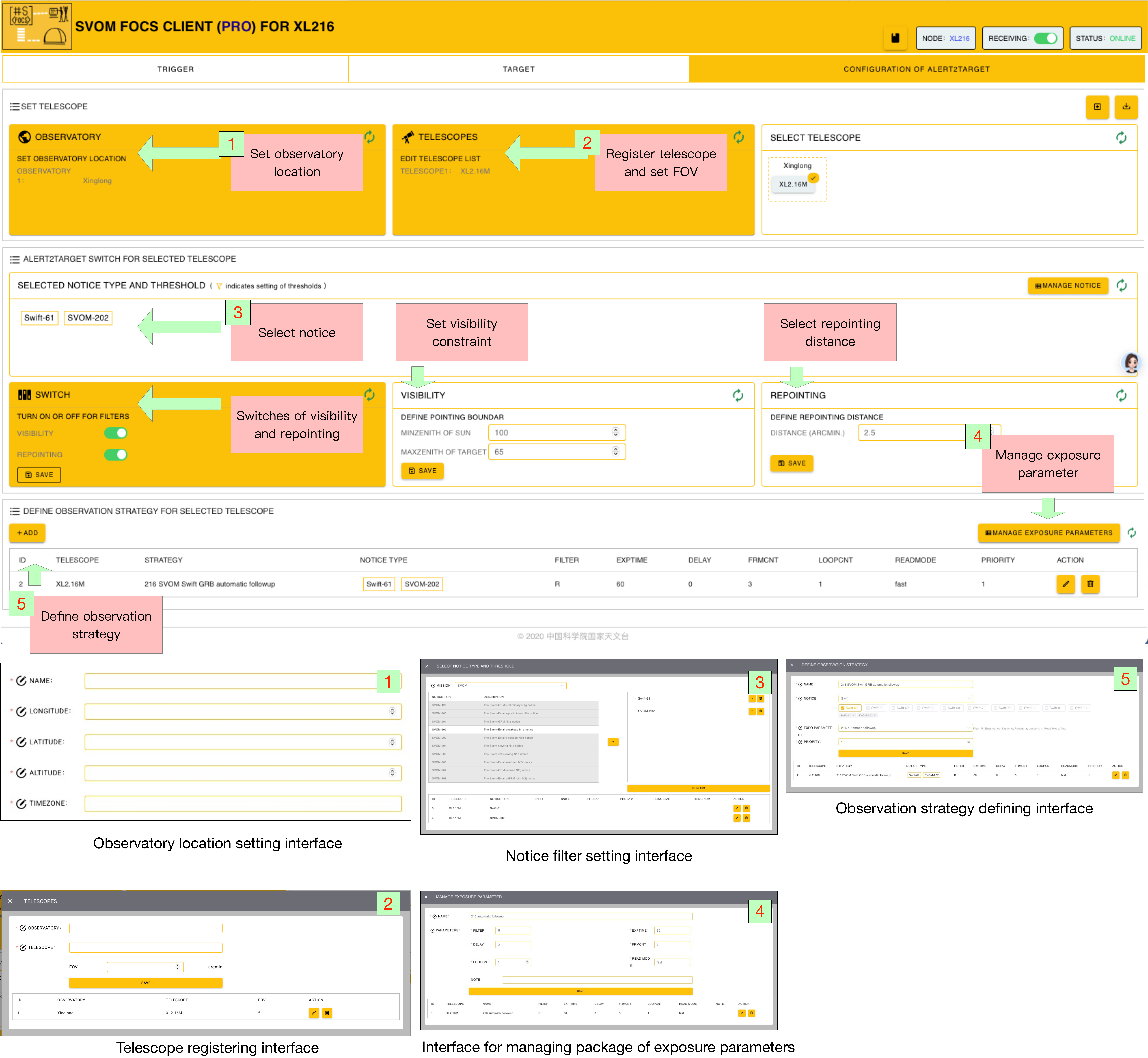}	
	\caption{An example of the interfaces of client for setting the follow-up observation planning strategies and parameter configurations. Function of each area is tagged, and configuration of each pop-up window is marked with number.} 
	\label{fig_client_config}
\end{figure*}

Figure \ref{fig_client_config} shows the client HMI for follow-up observation planning strategies and parameter configurations. The client HMI clearly demonstrates the process by which users define automatic follow-up observation strategies. First, users need to provide the client with the geographic location of the observatory, a list of telescopes, and the visible FOV for each telescope. Then, select specific alerts for each telescope (set alert filtering conditions). For example, users can select early alert types to achieve earlier observations, thereby detecting earlier changes in the GRB optical counterpart and increasing the chances of obtaining high-value scientific results. Users can also choose later-arriving alert types with higher certainty and more accurate coordinates to ensure the effective use of telescope observation time. For specific alert types, users can also set a signal-to-noise ratio (SNR) threshold, and only alerts with a SNR above the threshold will be used for follow-up observations.

Subsequently, the client provides users with two switches: (1) target visibility calculation tool switch; (2) telescope re-pointing necessity check tool switch. If the user turns on the target visibility calculation tool switch, the client will calculate whether the coordinates in the alert have an observable time window for the telescope. This tool requires the user to provide thresholds for solar altitude and target horizon altitude. If the user turns on the telescope re-pointing necessity check tool, the client will check whether the distance between the coordinates in the alert and the coordinates in the previous alert being observed by the telescope exceeds a given threshold (usually half the size of the telescope visible FOV). If the threshold is not exceeded, there is no need to rotate the telescope to point to the coordinates in the new alert; otherwise, it needs to be re-pointed.

In addition, the client allows users to pre-define a series of fixed combinations of exposure parameters for calling when formulating observation strategies for different types of alerts, without having to repeatedly define exposure parameters. Exposure parameters include band, exposure time, number of image frames, number of rounds, etc. Finally, when defining an observation strategy, the user needs to specify the alert type, select an exposure parameter combination for that type, and set the priority. Ultimately, after receiving an alert, the FOCS Alert2Target program will automatically formulate an observation plan based on the observation strategy.

\section{Alert2Target Software: The Core of GRB Automatic Follow-up Observation Planning}

\begin{figure}
	\centering 
	\includegraphics[width=0.35\textwidth]{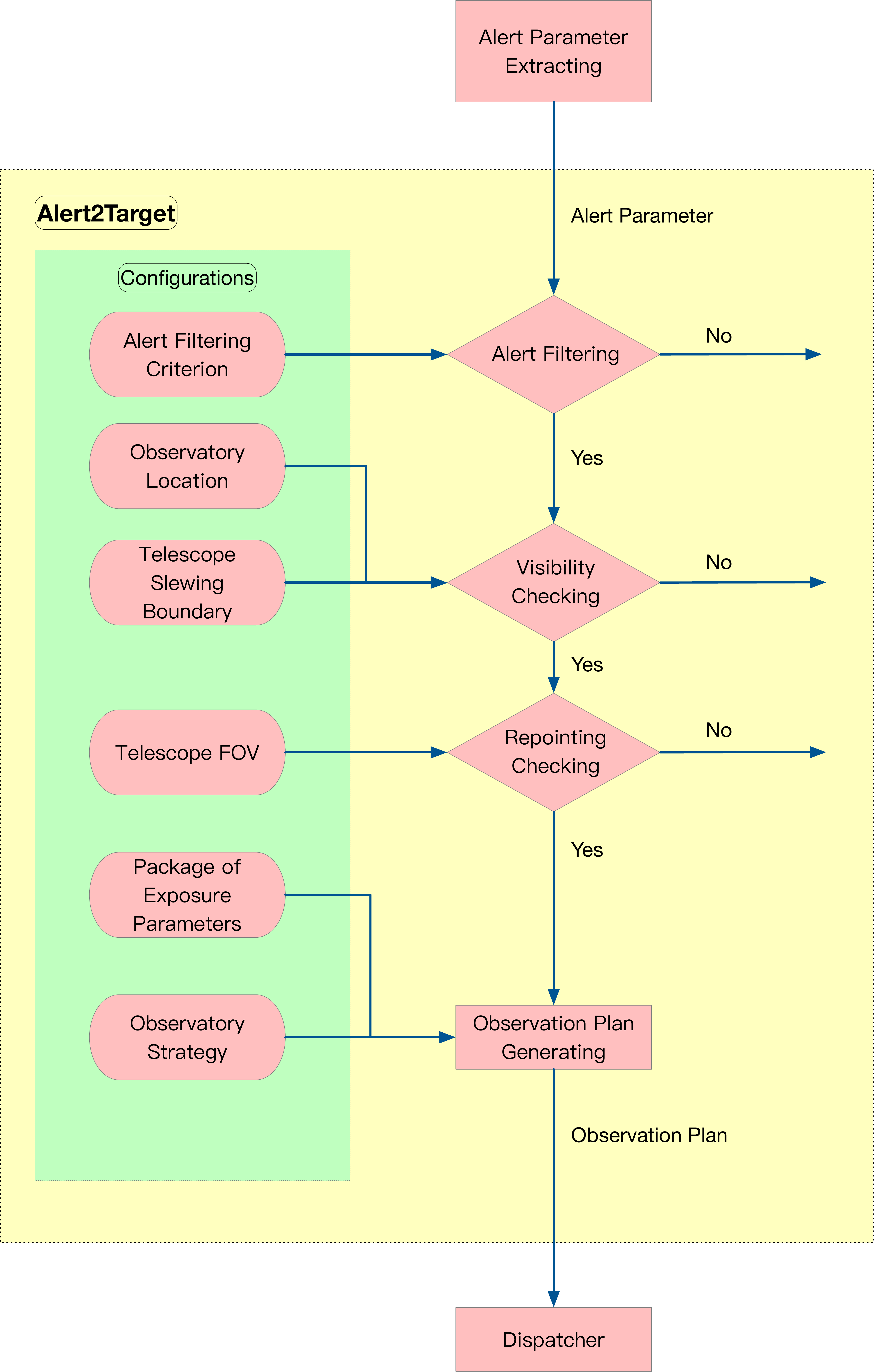}	
	\caption{The processing logic of Alert2Target, where the green box contents the pre-defined configurations and strategies.} 
	\label{fig_alert2target_scheduler}
\end{figure}

The Alert2Target software is a core component of the FOCS server and client business processes, responsible for implementing the GRB automatic follow-up observation planning functionality. The processing workflow is as follows: first, upon receiving an alert, the software filters the alerts based on predefined criteria, eliminating those that do not meet the conditions. Next, it calculates the visibility time window for the alert coordinates, taking into account the telescope’s geographical location, pointing range, or safety envelope, thereby filtering out alerts that do not meet visibility conditions. Third, it assesses the deviation between the alert coordinates and the current target coordinates. If the deviation exceeds a predefined threshold, a new observation plan is generated, prompting the telescope to point to the new coordinates; otherwise, the telescope continues with its current observation task. Finally, based on the type of alert, predefined observation strategies are matched for each telescope, resulting in the generation of detailed observation plans. The processing logic is depicted in Figure \ref{fig_alert2target_scheduler}.

\begin{figure}
	\centering 
	\includegraphics[width=0.48\textwidth]{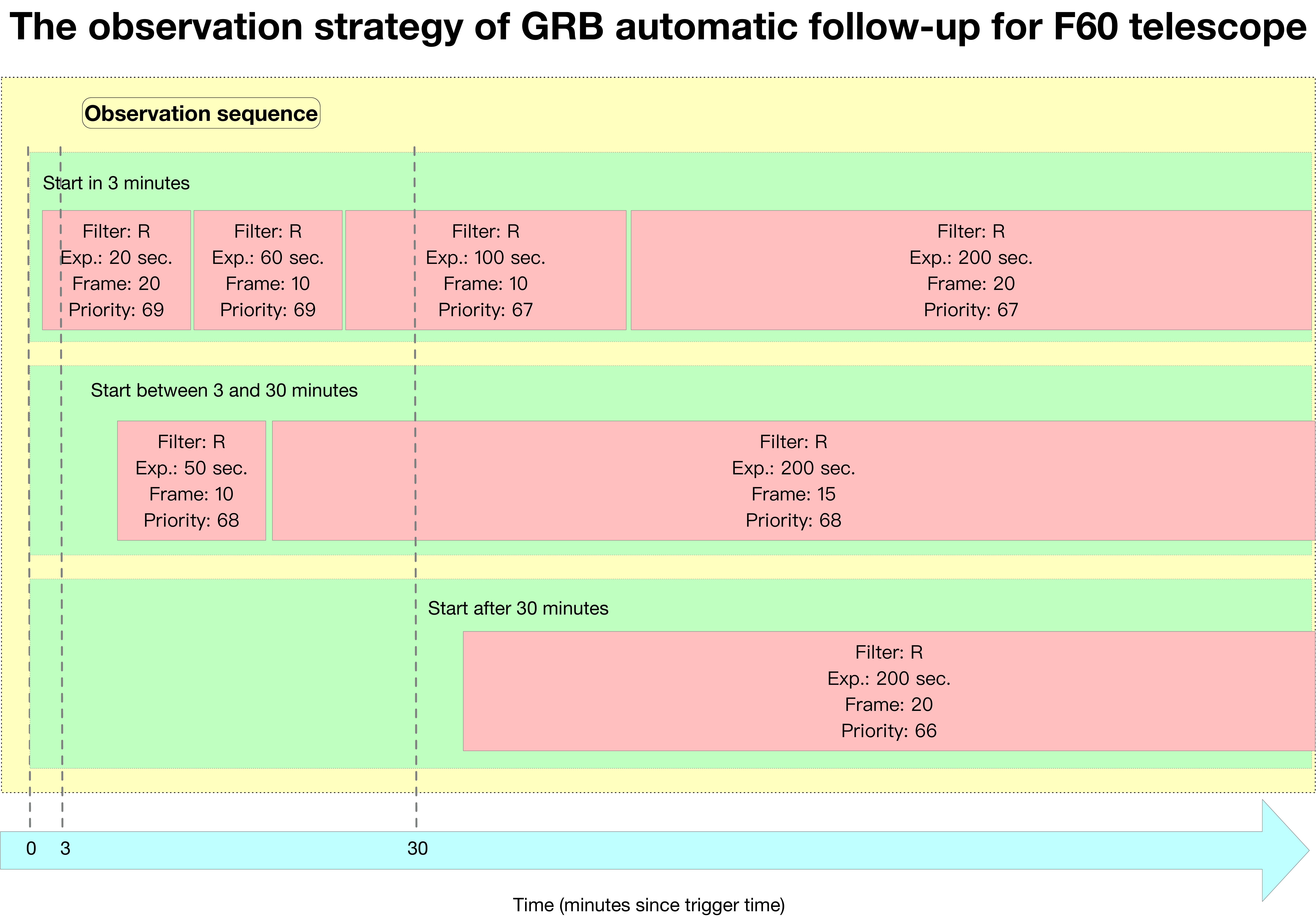}	
	\caption{The three types of temporal optimization strategies for automatic follow-up observations of SVOM GRB alerts adopted by the F60 telescope within the GWAC system. The red boxes content 
    the exposure parameters of single observation in the sequences.} 
	\label{fig_f60_obs_strategy}
\end{figure}

Initially designed to provide GRB automatic follow-up observation capabilities for telescopes, Alert2Target started as a simple backend Python program. However, as the number of users and the diversity of demands for FOCS services increased, Alert2Target evolved into an independent subsystem that includes a frontend HMI, a backend business processing program, and a database. This evolution has endowed FOCS with greater flexibility and usability. Users can not only utilize it for automatic follow-up observations of SVOM GRBs but also for planning automatic follow-up observations of alerts from multiple satellites (e.g. Swift, EP). Through meticulous design of exposure parameter combinations and priorities, complex observation organization can be achieved. Notably, the F60 telescope within the GWAC system has established a dynamic adaptive observation framework by integrating three types of temporal optimization strategies for automatic follow-up observations of SVOM GRB alerts, categorizing them based on the light curve characteristics of GRBs into three observation sequences according to the start time of observations:

\begin{itemize}
\item Within 3 minutes of the trigger: A short-medium-long exposure time sequence is employed to prioritize the acquisition of early light curve features and detect late variations;
\item 3-30 minute window: A medium-long exposure sequence optimizes the detection of light curve features and optical counterparts;
\item Beyond 30 minutes: A long exposure mode is utilized to focus on the search for optical counterparts.
\end{itemize}
This phased observation strategy, demonstrated in the Figure \ref{fig_f60_obs_strategy} effectively addresses the observational needs across different evolutionary stages of GRBs.

\section{Multi-Protocol Collaborative Communication Architecture: Ensuring Data Transmission in the FOCS System}

The FOCS system involves external communication with the CSCi and the ground telescope OCSs, as well as internal communication between the server and client. To meet the data transmission needs in various scenarios, FOCS employs a hybrid communication protocol stack.

Considering that astronomical observation sites are typically deployed in remote areas far from urban power grids, the network infrastructure is relatively weak, and communication links may experience high latency, low bandwidth, and intermittent disruptions, FOCS utilizes the MQTT protocol for real-time alert transmission from the server to distributed clients. The MQTT protocol is lightweight, efficient, and reliable, ensuring the dependable delivery of alerts even under extreme conditions. Its topic-based publish-subscribe mechanism achieves 1:N broadcast efficiency, while the Quality of Service (QoS) management ensures efficient transmission under network congestion. Additionally, the Last Will and Retained Message mechanisms provide state synchronization in the event of network disconnections.

For the uplink from the client to the server, FOCS employs the mature HTTP protocol. The HTTP protocol offers high reliability, cross-platform compatibility, and scalability, ensuring the complete and efficient transmission of non-real-time information or data, such as node authentication, status feedback, and data uploads (particularly large files of observational data).

\section{FOCS Release and Implementations}

After three major versions and numerous minor iterations, the FOCS service has achieved all the functionalities described in the previous sections and is now providing stable services. The FOCS client continuously improves its features through a repeated cycle of user usage, feedback, and updates, enhancing user convenience and enabling users to complete the vast majority of automatic follow-up observation tasks. 

Currently, the third version of the FOCS service has been launched, and the client software has been released simultaneously. Due to its ability to be defined and deployed anytime and anywhere, as well as its support for dynamic addition, deletion, and elastic scaling, the FOCS client has been deployed on twelve (sets of) telescopes in China and the United States, collectively forming the ground-based follow-up observation network for SVOM. As of January 2025, FOCS has coordinated these telescopes to complete dozens of follow-up observations and has issued 41 GCN (Gamma-ray Coordinates Network) circulars.

\begin{table*}
\begin{tabular}{l l l l l} 
 \hline
 Class & Telescope        & Location                    & Service of FOCS     & Telescope characteristics \\[5pt] 
 \hline
 A1    & CGFT$^{a}$       & Jilin Observatory, China    & observation plan    & automatic follow-up, \\
       &                  &                              &                    & feedback of observation status \\[5pt]
 \hline
 A1    & GWAC-A$^{b}$     & Xinglong Observatory, China  & observation plan    & automatic satellite-ground collaborative \\
       &                  &                              &                    & observation, feedback of observation status \\[5pt]
 \hline
 A1    & F60A/B, F50, F30$^{b}$ & Xinglong Observatory, China  & observation plan    & automatic follow-up, \\
       &                  &                              &                    & feedback of observation status \\[5pt]
 \hline
 A1    & WHO-1m$^{c}$           & Weihai Observatory, China    & observation plan    & automatic follow-up \\[5pt]
 \hline
 A2    & Mephisto-1.6m$^{d}$    & Lijiang Observatory, China   & observation plan    & semi-automatic follow-up$^{*}$\\[5pt]
 \hline
 A2    & Xinglong-2.16m$^{e}$   & Xinglong Observatory, China  & observation plan    & semi-automatic follow-up$^{*}$\\[5pt]
 \hline
 A3    & KAIT$^{f}$             & Lick Observatory, USA         & alert information   & automatic follow-up\\[5pt]
 \hline
 A3    & WFST$^{g}$             & Lenghu Observatory, China    & alert information   & automatic follow-up\\[5pt]
 \hline
 A4    & Lijiang-2.4m$^{h}$            & Lijiang Observatory, China   & alert information   & manual controlled follow-up\\[5pt]
 \hline
\end{tabular}
\caption{list of the telescopes and their link characteristics: 
\textit{a} refers to \citet{wu25}, 
\textit{b} refers to \citet{xin23},
\textit{c} refers to \citet{hu14},
\textit{d} refers to \citet{yang24},
\textit{e} refers to \citet{fan16},
\textit{f} refers to \citet{Filippenko01},
\textit{g} refers to \citet{wang23},
\textit{h} refers to \citet{fan15}. 
\textit{*}The "semi-automatic" indicates that the follow-up observation plan needs to be validated by the telescope operator, before starting the observation. }
\label{Table1}
\end{table*}

Based on the types of services utilized, these telescopes can be categorized into four classes: Class A1 and A2 telescopes use observation plans provided by FOCS, while Class A3 and A4 telescopes utilize alert information provided by FOCS.

To facilitate the use of FOCS services by more telescopes and users, we have integrated alert information from multiple astronomical satellites, including Swift, Fermi, and EP, based on SVOM GRB alerts, and standardized the output. Additionally, we have implemented a “software product layering” approach for the third version of the FOCS client software, customizing the software for different user groups and needs. Besides the professional version of the client, which includes complete functionality for the aforementioned twelve (sets of) telescopes, we also offer a simplified version of the client software. The simplified version provides alert information display, alert filtering, and various standardized alert information output interfaces, such as JSON files (local storage) and HTTP APIs (alert information query interface). The simplified client software is available for download 
(http://www.svom-gwacn.cn/doc/focs-client.zip). Non-SVOM core telescopes or non-professional astronomical users can download and use the FOCS client for GRB automatic follow-up observations.

\section{Performance Evaluation of the FOCS System}

The FOCS C/S architecture has achieved stable operation and high link reliability across the deployed telescopes. Deployment tests indicate that the FOCS system demonstrates good communication performance across different network environments:
\begin{itemize}
\item CGFT telescope (high-quality network, smaller message size): 97\% of alert messages are transmitted end-to-end within 3 seconds;
\item GWAC telescope (unstable network, larger message size): 90\% of observation plan messages have a transmission latency of less than 10 seconds;
\item Overall network message loss rate is close to zero (0\%).
\end{itemize}
These performance metrics validate the effectiveness of the hybrid protocol architecture in complex network environments, providing reliable support for real-time astronomical observation responses.

\section{Evolution of the FOCS System and Future Plans}

The ongoing development of the fourth-generation system (v4.0) focuses on the following key areas:

\paragraph{Multi-Satellite Collaborative Observation Module} By integrating the ToO observation interface of the SVOM satellite, we aim to achieve satellite-to-satellite (Swift, EP →  SVOM) collaborative observations.

\paragraph{Scientific Data Management System} Through the FOCS client API, we will implement associated storage of telescope scientific data and cross-device multidimensional retrieval.

These upgrades will provide more robust technical support for time-domain astronomy research.

\section{Conclusion}

The Sino-French SVOM mission marks a transformative advancement in the study of GRBs and other variable cosmic phenomena, leveraging cutting-edge space-based and ground-based instrumentation. The successful implementation of the FOCS system, with its robust communication performance and hybrid protocol architecture, has demonstrated its capability to support real-time astronomical observations across diverse network environments. As the system evolves with the integration of multi-satellite collaborative observation modules and enhanced scientific data management capabilities, SVOM is poised to significantly contribute to time-domain astronomy research.

\section*{Acknowledgements}

The authors are thankful for support from the National Key R\&D Program of China (grant No.2024YFA1611700, 2024YFA1611702, 2023YFA1608304).
This work is also supported by the Supported by the Strategic Priority Research Program of the Chinese Academy of Sciences (grant No. XDB0550101) and the National Natural Science Foundation of China (grant No. 11903054, 12273054).
We thank the staffs of the Xinglong-2.16m telescope, and Jilin, Lijiang, Weihai observatories for important assistants during the deployment of the FOCS clients. This work was partially supported by the  Open  Project  Program  of  the  Key  Laboratory  of  Optical  Astronomy,  National Astronomical Observatories, Chinese Academy of Sciences.”


\begin{thebibliography}{00}



\bibitem[Atteia (2022)]{Atteia22}
  Atteia, J-L., Cordier, B., Wei, J.,
  The SVOM Mission,
  arXiv:2203.10962v1,
  2022.

\bibitem[Fan (2015)]{fan15}
  Fan, Y., Bai, J., Zhang, J., et al.,
  Rapid instrument exchanging system for the Cassegrain focus of the Lijiang 2.4-m Telescope,
  Research in Astronomy and Astrophysics,
  Volume 15, Issue 6, article id. 918,
  2015.
  
\bibitem[Fan (2016)]{fan16}
  Fan, Z., Wang, H., Jiang, X., et al.,
  The Xinglong 2.16-m Telescope: Current Instruments and Scientific Projects,
  Publications of the Astronomical Society of the Pacific,
  Volume 128, Issue 969, pp. 115005,
  2016.

\bibitem[Filippenko (2001)]{Filippenko01}
  Filippenko, A., Li, W., Treffers, R., Modjaz, M.,
  The Lick Observatory Supernova Search with the Katzman Automatic Imaging Telescope,
  Small Telescope Astronomy on Global Scales, Astronomical Society of the Pacific,
  ISBN: 1-58381-084-6, 2001., p.121,
  2016.

\bibitem[Hu (2014)]{hu14}
  Hu, S., Han, S., Guo, D., Du, J.,
  The photometric system of the One-meter Telescope at Weihai Observatory of Shandong University,
  Research in Astronomy and Astrophysics,
  Volume 14, Issue 6, article id. 719-732,
  2014.

\bibitem[Wang (2023)]{wang23}
  Wang, T., Liu, G., Cai, Z., et al.,
  Science with the 2.5-meter Wide Field Survey Telescope (WFST),
  Science China Physics, Mechanics \& Astronomy,
  Volume 66, Issue 10, article id.109512,
  2023.

\bibitem[Wu (2025)]{wu25}
  Wu, C., et al.,
  GRB 240825A: Early Reverse Shock and Its Physical Implications,
  submitted,
  2025.

\bibitem[Xin (2023)]{xin23}
  Xin, L., Han, X., Li, H., et al.,
  Prompt-to-afterglow transition of optical emission in a long gamma-ray burst consistent with a fireball,
  Nature Astronomy,
  Volume 7, p. 724-730,
  2023.

\bibitem[Yang (2024)]{yang24}
  Yang, Y., Jiu, X., Pan, Y., et al.,
  Multiband Simultaneous Photometry of Type II SN 2023ixf with Mephisto and the Twin 50 cm Telescopes,
  The Astrophysical Journal,
  Volume 969, Issue 2, id.126, 10 pp,
  2024.
  
\end{thebibliography}


\end{document}